\documentclass[aps, prd, twocolumn, superscriptaddress, preprintnumbers, nofootinbib, longbibliography]{revtex4-1}

\usepackage{amsmath}	% AMS Math Package
\usepackage{amsthm}		% Theorem Formatting
\usepackage{amssymb}	% Math symbols such as \mathbb
\usepackage{eufrak}
\usepackage{datetime}
\usepackage{graphicx}
\usepackage{color}
\usepackage{verbatim}
\usepackage{bm}

\usepackage[colorlinks=true, citecolor=midblue, linkcolor=midblue, urlcolor=midblue]{hyperref}

\usepackage{setspace}

\settimeformat{ampmtime}

\definecolor{grey}{rgb}{0.4,0.4,0.4}
\definecolor{dullmagenta}{rgb}{0.4,0,0.4}
\definecolor{darkblue}{rgb}{0,0,0.4}
\definecolor{midblue}{rgb}{0,0,0.5}
\definecolor{midred}{rgb}{0.5,0,0}
\definecolor{orange}{rgb}{1,0.5,0}
\definecolor{lightbrown}{rgb}{0.75,0.5,0.25}
\definecolor{tan}{cmyk}{0.14,0.42,0.56,0}
\definecolor{djunglegreen}{cmyk}{0.99,0,0.52,0}
\definecolor{lightgreen}{rgb}{0,1,0}
\definecolor{olivegreen}{cmyk}{0.64,0,0.95,0.40}
\definecolor{midgreen}{rgb}{0.0,0.675,0.0}
\definecolor{darkgreen}{rgb}{0,0.5,0}

\newcommand{\vs}{\vspace}

\renewcommand{\.}{\hspace{0.5mm}}

\newcommand{\ra}{\ensuremath{\rightarrow}}

\newcommand{\Prm}{\ensuremath{\mathrm{P}}}

\newcommand{\crm}{\ensuremath{\mathrm{c}}}
\newcommand{\drm}{\ensuremath{\mathrm{d}}}
\newcommand{\erm}{\ensuremath{\mathrm{e}}}

\newcommand{\grm}{\ensuremath{\mathrm{g}}}
\newcommand{\hrm}{\ensuremath{\mathrm{h}}}
\newcommand{\irm}{\ensuremath{\mathrm{i}}}

\newcommand{\qrm}{\ensuremath{\mathrm{q}}}

\newcommand{\srm}{\ensuremath{\mathrm{s}}}

\newcommand{\Ocal}{\ensuremath{\mathcal{O}}}

\newcommand{\eg}{e.g.}
\newcommand{\ie}{i.e.}

\newcommand{\cf}{c.f.}

\settimeformat{ampmtime}

\begin{document}
%%%%%%%%%%%%%%%%%%%%%%%%%%%%%%%%%%%%%%%%%%%%%%%%%%%%%%%%
\title{Primordial Black Holes from Confinement}

%%%%%%%%%%%%%%%%%%%%%%%%%%%%%%%%%%%%%%%%%%%%%%%%%%%%%%%%
\author{Gia Dvali}
\affiliation{
	Arnold Sommerfeld Center,
	Ludwig-Maximilians-Universit{\"a}t,
	Theresienstra{\ss}e 37,
	80333 M{\"u}nchen,
	Germany}
\affiliation{
	Max-Planck-Institut f{\"u}r Physik,
	F{\"o}hringer Ring 6,
	80805 M{\"u}nchen,
	Germany}

\author{Florian K{\"u}hnel}
\email{Florian.Kuehnel@physik.uni-muenchen.de}
\affiliation{
	Arnold Sommerfeld Center,
	Ludwig-Maximilians-Universit{\"a}t,
	Theresienstra{\ss}e 37,
	80333 M{\"u}nchen,
	Germany}

\author{Michael Zantedeschi}
\email{Michael.Zantedeschi1@physik.uni-muenchen.de}
\affiliation{
	Arnold Sommerfeld Center,
	Ludwig-Maximilians-Universit{\"a}t,
	Theresienstra{\ss}e 37,
	80333 M{\"u}nchen,
	Germany}
\affiliation{
	Max-Planck-Institut f{\"u}r Physik,
	F{\"o}hringer Ring 6,
	80805 M{\"u}nchen,
	Germany}

%%%%%%%%%%%%%%%%%%%%%%%%%%%%%%%%%%%%%%%%%%%%%%%%%%%%%%%%
\date{\formatdate{\day}{\month}{\year}, \currenttime}

%%%%%%%%%%%%%%%%%%%%%%%%%%%%%%%%%%%%%%%%%%%%%%%%%%%%%%%%
\begin{abstract}
A mechanism for the formation of primordial black holes is proposed. Here, heavy quarks of a confining gauge theory produced by de Sitter fluctuations are pushed apart by inflation and get confined after horizon re-entry. The large amount of energy stored in the colour flux tubes connecting the quark pair leads to black-hole formation. These are much lighter and can be of higher spin than those produced by standard collapse of horizon-size inflationary overdensities. Other difficulties exhibited by such mechanisms are also avoided. Phenomenological features of the new mechanism are discussed as well as accounting for both the entirety of the dark matter and  the supermassive black holes in the galactic centres. Under proper conditions, the mechanism can be realised in a generic confinement theory, including ordinary QCD. We discuss a possible string-theoretic realisation via $D$-branes. Interestingly, for conservative values of the string scale, the produced gravity waves are within the range of recent NANOGrav data. Simple generalisations of the mechanism allow for the existence of a significant scalar component of gravity waves with distinct observational signatures.

\end{abstract}


\maketitle

%%%%%%%%%%%%%%%%%%%%%%%%%%%%%%%%%%%%%%%%%%%%%%%%%%%%%%%%%%%%%%
\section{Introduction}
\label{sec:Introduction}

The milestone discovery of black-hole mergers by LIGO/Virgo \cite{TheLIGOScientific:2016pea}, with recently several dozens more \cite{Abbott:2020jks}, has fostered interest in them. As more of these merging objects are being found with masses at the boundary or even outside of what can be expected from stellar collapse, the following question arises: Could these black holes have been produced primordially?

There are several formation mechanisms for these primordial black holes (PBHs). All of them require large overdensities, which collapse to black holes if above a critical threshold. In most scenarios, these overdensities are of inflationary origin, and the collapse to PBHs occurs when the associated scale re-enters the Hubble horizon. Besides, there are other non-inflationary scenarios for PBH formation{\;---\;}for instance where the inhomogeneities arise from first-order phase transitions, bubble collisions, and the collapse of cosmic strings, necklaces, domain walls or non-standard vacua (see Ref.~\cite{Carr:2020xqk} for references). Despite plentitude of attempts, the understanding of viable mechanisms for the production of PBHs remains one of the most interesting and challenging questions of particle physics and cosmology. 

In this work we present a novel mechanism for producing PBHs, which is rather natural within the framework of confining theories, such as, asymptotically-free QCD-like gauge theories with massive quarks. Key to this mechanism is the dilution of heavy quarks produced during inflation which get confined by QCD flux tubes after horizon re-entry. PBH formation is then induced by the very large amount of energy which is stored in the string connecting the quark pair. This effect shares some similarity with the production of black holes via the collapse of cosmic strings attached to monopoles \cite{matsuda}. Of course, there are fundamental differences. Firstly, quarks are produced as perturbative states during inflation. Secondly, their confinement by QCD flux tubes is guaranteed by the asymptotic freedom of the theory. This makes an implementation of our mechanism possible even within the Standard Model QCD, provided certain conditions, to be specified below, in the early epoch are met.

We also speculate about the string-theoretic realisations of this mechanism. In this case, the r{\^o}le of heavy quarks is played by compact $D$-branes produced by de Sitter fluctuations during brane inflation. The r{\^o}le of the QCD flux tubes connecting them are played by $D$-strings which form after graceful exit. We estimate that for conservative values of string-theoretic parameters, the gravitational waves produced during the confinement process is in interesting range for NANOGrav data \cite{NANOGrav:2020bcs}. 

Simple generalisations of our scenario can result into the existence of a scalar component of the emitted gravitational waves from PBH formation. This can lead to potentially interesting observational signatures such as the existence of significant low multipoles as well as the dependence of detected power of the gravity waves on the isotope composition of the detector. In fact, as indicated by NANOGrav observations, current observational sensitivity does not exclude the possibility of a further non-quadrupolar contribution to the signal. If this possibility is verified by future observations, it could serve as a hint of such a non-standard gravitational-wave contributions.

This paper is structured as follows. In Sec.~\ref{sec:Mechanism} the new mechanism is introduced and its dynamics are described. Sec.~\ref{sec:Dark-Matter} elaborates on its cosmological relevance for the production of PBH dark matter as well as seeds for the supermassive black holes in galactic centres. Furthermore, the viability of our scenario for real QCD as well as in string-theoretic contexts is addressed. We point out how to obtain a scalar component of the emitted gravitational waves from PBH formation. We close with a Discussion and Outlook in Sec.~\ref{sec:Discussion-and-Outlook}. 

Throughout this work we utilise units in which $M_{\Prm} = c = \hbar = 1$, unless explicitly stated for clarity.

%%%%%%%%%%%%%%%%%%%%%%%%%%%%%%%%%%%%%%%%%%%%%%%%%%%%%%%%%%%%%%
\section{Mechanism}
\label{sec:Mechanism}

The key ingredient of our scenario is a QCD-like gauge theory with massive quarks. The term ``quark" is very general and implies an arbitrary coloured state onto which the gauge flux tube can end. An important requirement is that all quarks must be heavier than the corresponding QCD scale, $\Lambda_{\crm}$, throughout the epoch of interest. Therefore the string does not fragment. This epoch covers the last $60$ or so $e$-folds of the inflationary period, graceful exit, reheating and the part of radiation domination during which the black holes are formed. In the later epoch the balance between the quark masses and the QCD scale can change and some of the quarks can become lighter than $\Lambda_{\crm}$. This later variation shall not affect the outcome of PBH formation. Since, generically the expectation values of the fields change throughout cosmology, the class of possibility is rather wide and can include even ``our" QCD.
 
In the simplest scenario, we can assume that during inflation QCD was not in a confining phase. That is, $\Lambda_{\crm}$ evaluated at that time, was lower than the value of the inflationary Hubble parameter $H_{\irm}$. Under such circumstances, the quark/anti-quark pairs produced by de Sitter fluctuations, will be freely inflated away. 

In the opposite case, if $\Lambda_{\crm} > H_{\irm}$, quarks will be connected by the QCD flux tubes from the moment of their creation. This suppresses their nucleation probability because of additional contributions from the flux tubes.
    
The proposed scenario then works as follows. Let us consider a quark/anti-quark pair, which is produced during the inflationary period. Let their mass $m_{\qrm}$ be bigger but somewhat comparable to $H_{\irm}$ and assume that this pair is first inflated outside of the Hubble horizon and comes in causal contact again during radiation domination when it gets confined by strings of tension $\mu = \Lambda_{\crm}^{2}$. Then, the energy stored in the confining string is 
\begin{equation}
	\label{eq:energystring}
    E
        \sim
                    \Lambda_{\crm}^{2}\,t
                    \; ,
\end{equation}
where $t$ is the time of horizon re-entry of the pair.

\begin{figure}[t]
  \centering
  \includegraphics[width=0.45\textwidth]{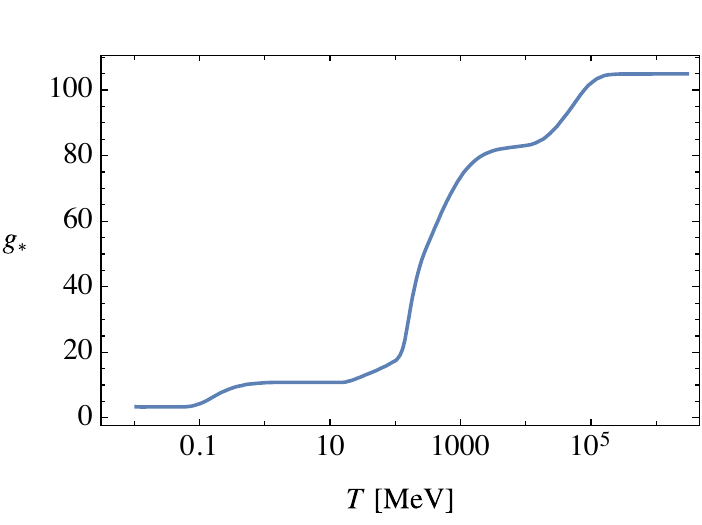}
  \caption{Number of relativistic degrees of freedom as a function of temperature 
      (adopted from Ref.~\cite{Carr:2019kxo}).}
  \label{fig:gt}
\end{figure}

Once within causal contact, the two quarks start moving towards each other with acceleration proportional to $\Lambda_{\crm}^{2} / m_{\qrm}$, becoming quickly ultra-relativistic. Due to the large amount of energy stored in the confining string, the system's Schwarzschild radius $R_{\grm} \sim E$ is much bigger than the string width $\Lambda_{\crm}^{-1}$. This implies the condition
\begin{equation}
    \label{eq:CCC}
    \frac{\Lambda_{\crm}^{2}}{\sqrt{g_{*}( T )\,}T^{2}}
        \gg 
                    \frac{ 1 }{ \Lambda_{\crm} }
                    \; ,
\end{equation}
resulting in the formation of a PBH of mass $M_{\rm PBH}$,
\begin{equation}
\label{eq:gravmass1}
    M_{\rm PBH}( t )
        \sim
                    E( t )
        \sim
                    \Lambda_{\crm}^{2}\,t
        \gg
                    \Lambda_{\crm}^{-1}
                    \; ,
\end{equation}
which can readily be expressed in terms of the temperature $T$ at the time of formation:
\begin{equation}
\label{eq:ttoT}
    M_{\rm PBH}( T )
        \approx
                    \frac{ 3\sqrt{5\,}\,\Lambda_{\crm}^{2} }
                    { 4 \pi ^{3/2} \sqrt{g_{*}( T )\,}\,T^{2} }
                    \; .
\end{equation}
Here, $g_{*}( T )$ is the number of relativistic degrees of freedom at temperature $T$ (depicted in Fig.~\ref{fig:gt}). 

\begin{figure}[t]
  \centering
  \includegraphics[width=0.48\textwidth]{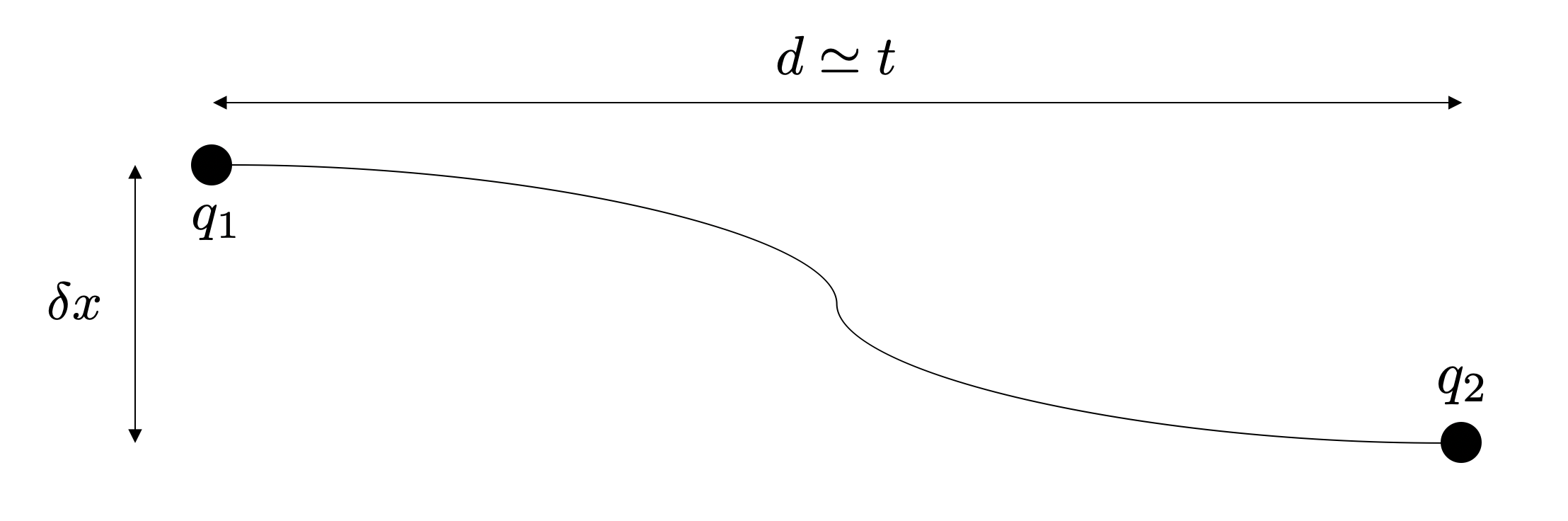}
  \caption{Sketch of the initial configuration at collapse, 
      for which $\delta x \ll d$.
      }
  \label{fig:stringbh}
\end{figure}

The PBH mass in Eq.~\eqref{eq:gravmass1} should be compared with the one obtained from the collapse of horizon-size overdensities \cite{Hawking:1971ei, 1967SvA....10..602Z, Carr:1975qj}, in which case we have
\begin{equation}
    \widetilde{M}_{\rm PBH}( t )
        \approx
                    \gamma\,t
                    \; ,
\end{equation}
with $\gamma \sim \Ocal( 1 )$.\footnote{Here, for simplicity, we neglect the effect of critical collapse (\cf~Refs.~\cite{Choptuik:1992jv, Koike:1995jm, Niemeyer:1997mt, Evans:1994pj, Gundlach:1999cu, Musco:2012au, Kuhnel:2015vtw}), which leads to a PBH mass spectrum.} Hence, $\widetilde{M}_{\rm PBH}$ is much larger than that of our proposed mechanism; their ratio being
\begin{equation}
    \frac{ M_{\rm PBH} }{ \widetilde{M}_{\rm PBH} }
        \approx
                    \frac{ \Lambda_{\crm}^{2} }{ \gamma }
        \ll
                    1
                    \; .
\end{equation}

Before assessing under which conditions our mechanism is able to account for $100\%$ of PBH dark matter, let us discuss the spin of the resulting black holes.

%%%%%%%%%%%%%%%%%%%%%%%%%%%%%%%%%%%%%%%%%%%%%%%%%%%%%%%%%%%%%%
\subsection*{Spin}
\label{sec:Spin}

During the inflationary dilution, the string zero mode undergoes a Brownian motion induced by de Sitter quantum fluctuations which behave very much as the thermal effects described in Sec.~\ref{sec:Dark-Matter}. The fluctuation per Hubble time $H_{\irm}^{-1}$ is of order $H_{\irm}$. Therefore the string's deviation from straightness is of order
\begin{equation}
    \label{eq:brownian}
    \delta x
        \simeq
                    \sqrt{N_{\erm}\,}\,H_{\irm}^{-1}
                    \; ,
\end{equation}
where $N_{\erm}$ is the number of $e$-folds from the string-pair nucleation time to the end of inflation, and is given by
\begin{equation}
    N_{\erm}
     =
                    \frac{ 1 }{ 2 }\.
                    \log\!
                    \left(
                        \frac{ M_{\rm PBH}\.H_{\irm} }{ \Lambda_{\crm}^{2} }
                    \right)
                    .
\end{equation}
As the quark/anti-quark pair collapses in the radiation-dominated epoch{\;---\;}its motion being ultra-relativistic{\;---\;}Eq.~\eqref{eq:brownian} is a good estimate for the impact parameter at black-hole formation (see Fig.~\ref{fig:stringbh} for a sketch of the initial collapsing configuration). In the following we assume this to be the only mechanism sourcing angular momentum for PBHs. Therefore the dimensionless Kerr spin parameter $a_{\rm PBH}$ at formation is well-approximated by
\begin{equation}
\label{eq:apbh}
    a_{\rm PBH}
        \simeq
                    \frac{ \delta x }{ R_{\rm g} }
        \simeq
                    \frac{ 1 }{ H_{\irm}\.M_{\rm PBH} }\.
                    \log\!
                    \left(
                    \frac{ H_{\irm}\,M_{\rm PBH} }{ \Lambda_{\crm}^{2} }
                    \right)^{\!1/2}
                    ,
\end{equation}
where numerical factors of order one have been neglected.

Notice that the condition (\ref{eq:CCC}) will translate as a constraint $a_{\rm PBH} < \sqrt{ N_{\erm}}\.\Lambda_{\crm} / H_{\irm}$, where $\Lambda_{\crm}$ is the string scale during the re-entry. Therefore, one has to pay a special attention to the scenarios in which this quantity changes significantly relative to its inflationary value, as this can constrain the impact parameter significantly. In the simplest case when $\Lambda_{\crm} \sim H_{\irm} \sim R_{\grm}^{-1}$ throughout the process, Eq.~\eqref{eq:apbh} would result into a near-maximal spin.

\begin{figure}[t]
  \centering
  \includegraphics[width=0.48\textwidth]{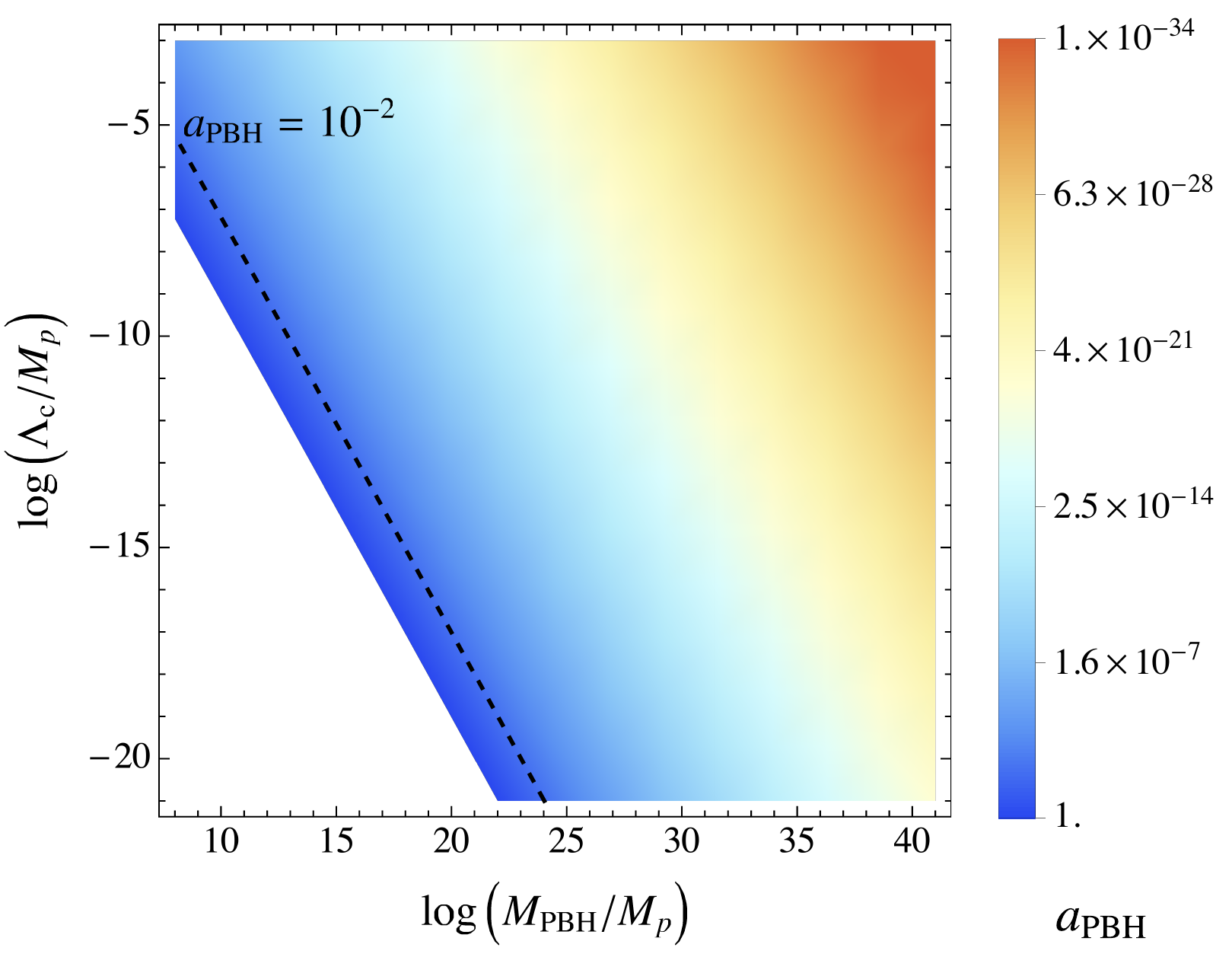}
  \caption{Dimensionless spin $a_{\rm PBH}$ at formation time as a function of $M_{\rm PBH}$ 
      and $\Lambda_{\crm}$ according to Eq.~\eqref{eq:apbh}. 
      The case $\Lambda_{\crm} \gtrsim H_{\irm}$ is considered.
      }
  \label{fig:spin}
\end{figure}
Equation~\eqref{eq:apbh} is depicted in Fig.~\ref{fig:spin}, where the case $\Lambda_{\crm} \gtrsim H_{\irm}$ is assumed. The white region corresponds to $a_{\rm PBH} > 1$, in which case PBH formation is unclear. In fact, in this case the impact parameter is bigger than the corresponding configuration's Schwarzschild radius. Consequently, the system will undergo rotations while radiating, making more accurate studies including back-reaction necessary. These are beyond the scope of this work.

Depending on the confinement scale $\Lambda_{\crm}$, highly-rotating black holes can be produced. For low values of mass and high confining scales (see Fig.~\ref{fig:spin}), it is even possible to produce {\it maximally}-spinning PBHs accounting for $f_{\rm PBH} = 1$ and being consistent with structure-formation constraints which demand the presence of dark matter at latest around redshift $\Ocal( 10^{4} )$ corresponding to $T \sim \Ocal( 10 )\,{\rm eV}$. Of course, isolated events of black-hole formation via confinement could still have taken place at much later time, and might occur even today, with potentially detectable signatures.

It is clear from Eq.~\eqref{eq:apbh} that for higher values of $H_{\irm}$ (and thereby higher confinement scales in the case $\Lambda_{\crm} \gtrsim H_{\irm}$, c.f.~Fig.~\ref{fig:spin}), heavier black holes will not be spinning significantly. Indeed, any sizable spin of black holes of mass $M_{\rm PBH} \sim M_{\odot} \sim 10^{38}\,M_{\Prm}$ cannot be caused by inflationary fluctuations as being clear from Fig.~\ref{fig:spin}. However, in this region PBHs can start spinning due to accretion \cite{de2020evolution}.

Finally, the window between $10^{4}\,\grm$ and $10^{10}\,\grm$, as discussed in Sec.~\ref{sec:Evaporation-Constraints}, is an interesting region in which higher inflationary scales can contribute to the PBH spin, therefore making it very attractive from an observational point of view. In general, probing PBH spins in the mass range close to the blue-coloured area of Fig.~\ref{fig:spin} could not only serve as a check of the proposed mechanism but also as an indirect measurement of the inflationary scale $H_{\irm}$.

The above discussion underlines that spin may be used as a discriminator between PBHs which form from confinement versus the standard mechanism via overdensities. In fact, while the former can give maximally-rotating black holes, $a_{\rm PBH} \sim 1$, the latter has an upper bound on the spin at formation $a_{\rm PBH}^{\rm overdense} \lesssim 10^{-1}$ as discussed in Ref.~\cite{Chiba:2017rvs} (see also Refs.~\cite{Mirbabayi:2019uph, DeLuca:2019buf}, and Ref.~\cite{Harada:2017fjm} for producing larger spins in the matter-dominated era). Note that accretion cannot help to significantly increase the PBH spin as long as $M_{\rm PBH} \ll M_{\odot}$ (see Ref.~\cite{de2020evolution}).

%%%%%%%%%%%%%%%%%%%%%%%%%%%%%%%%%%%%%%%%%%%%%%%%%%%%%%%%%%%%%%
\section{Dark Matter}
\label{sec:Dark-Matter}

Due to the wide range of PBH masses produced by the confinement mechanism, it is natural to ask whether it can form an order-one fraction of the dark matter. Our mechanism has some similarities with the one proposed in Refs.~\cite{Dengvilenkin, deng2017primordial2}. There, bubbles produced during the inflationary period grow exponentially due to the cosmological expansion, and then eventually re-collapse during the radiation-dominated era. In our scenario, as shown below, no fine-tuning of the production rate of quarks during the inflationary period is needed in order to account for the entirety of the dark matter.

The quark mass $m_{\qrm}$ can be either higher or lower than the inflationary Hubble scale $H_{\irm}$. In the former case nucleation processes are exponentially suppressed, similar to the dual monopole-case. In fact, objects heavier than the Hubble scale can be produced due to de Sitter Gibbons-Hawking effective temperature $T_{\hrm} = H_{\irm} / 2 \pi$ \cite{gibbons1977cosmological}. Their nucleation rate scales as \cite{basu1991quantum} 
\begin{equation}
    \label{eq:prob}
    \lambda
        \propto
                    e^{-B}
                    \; .
\end{equation}
In the case of quarks $B = m_{\qrm} / T_{\hrm} \equiv B_{\qrm}$ (see Ref.~\cite{basu1991quantum}). Either the confinement scale is higher than the inflationary Hubble parameter, implying that the quarks nucleate already in the confined phase, or the confinement takes place at lower temperatures later in the history of the Universe. In the former case, also the string nucleation probability contributes to Eq.~\eqref{eq:prob}, namely $B = B_{\qrm} + 2\.\Lambda_{\crm}^{2}\.R_{\srm} / T_{\hrm}$, where $R_{\srm}$ corresponds to the typical quark distance (and $R_{\srm}\geq H_{\irm}^{-1}$ is necessary in order to avoid immediate collapse). It can be shown that taking string-gravity effects into account, leaves the exponential suppression unaltered \cite{basu1991quantum}. Moreover, we need
\begin{align}
\label{eq:assumptscales}
    \Lambda_{\crm}
        \lesssim
                    m_{\qrm}
\end{align}
in order to avoid tunneling processes of the cosmo-logically-long strings into quarks-pairs upon horizon re-entry later in the Universe history. In fact, these processes have a probability per unit-length per unit-time given by $P \propto \exp( -\pi\,m_{\qrm}^{2} / \Lambda_{\crm}^{2} )$ (see Ref.~\cite{Vilenkin:2000jqa}).

The calculation of Eq.~\eqref{eq:prob} relies on the validity of the semi-classical approximation. Namely it will hold true as long as the quark masses $m_{\qrm}$ (and if confined during inflation, the confinement scale scale $\Lambda_{\crm}$) is bigger than the inflationary Hubble scale. Indeed we expect the heavy quarks not to be totally inflated away. In turn this means that we are interested in scenarios where $m_{\qrm}( \Lambda_{\crm} ) \gtrsim H_{\irm}$ implying that the dimensionless prefactor in Eq.~\eqref{eq:prob} is of order one.

Independently of the nucleation scenario it is possible to compute the PBH dark-matter fraction, being defined as
\begin{equation}
    \label{eq:fPBH-1}
    f_{\rm PBH}
        \equiv
                    \frac{ \rho_{\rm PBH}( t ) }{ \rho_{\rm CDM}( t ) }
        =
                    \frac{ 1 }{ \rho_{\rm CDM}( t ) }
                    \int \drm M_{\rm PBH}\;
                    \frac{ \drm\mspace{1mu} n_{\rm PBH}( t ) }
                    { \drm\mspace{-1mu} \ln M_{\rm PBH} }
                    \; ,
                    \vphantom{_{_{_{_{_{_{_{_{\phantom{\big|}}}}}}}}}}
\end{equation}
wherein $\rho_{\rm CDM}$ is the energy density of cold dark matter (CDM). Assuming that the collapse takes place in a radiation-dominated universe, it can be expressed as 
\begin{equation}
    \label{eq:rhocdm}
    \rho_{\rm CDM}( t )
        \approx
                    \frac{ 3 }{ 16 \pi\,t^{3/2} }
                    \frac{ 1 }{ M_{\rm eq}^{1/2} }
                    \; ,
\end{equation}
with $M_{\rm eq} = t_{\rm eq} = 2.8 \times 10^{17}\,M_{\odot} \sim 10^{55}$, corresponding to the cosmological horizon mass at the time $t_{\rm eq}$ of matter-radiation equality. The number density of PBHs redshifts in exactly the same way as $\rho_{\rm CDM}$:
\begin{equation}
    \label{eq:nPBH}
    \frac{ \drm\mspace{1mu} n_{\rm PBH}( t ) }
    { \drm\mspace{-1mu} \ln M_{\rm PBH} }
        \approx
                    \lambda \Lambda_{\crm}^{3}\,
                    \frac{ 1 }{ M_{\rm PBH}^{3/2} }\,
                    \frac{ 1 }{ t^{3/2} }
                    \; ,
\end{equation}
where a constant $\lambda$ given by Eq.~\eqref{eq:prob} was assumed through inflation (as discussed below Eq.~\eqref{eq:prob} the prefactor is of order-one). Combining Eqs.~(\ref{eq:rhocdm}, \ref{eq:nPBH}), we find for the present PBH dark-matter fraction, Eq.~\eqref{eq:fPBH-1},
\begin{equation}
\label{eq:fpbh}
    f_{\rm PBH}
        \approx
                    \frac{32 \pi}{3}\,
                    \lambda \Lambda_{\crm}^{3}
                    \left(
                        \frac{ M_{\rm PBH} }{ M_{\rm eq} }
                    \right)^{\!- 1/2}
                    \; .
\end{equation}

\begin{figure}[t]
  \centering
  \includegraphics[width=0.48\textwidth]{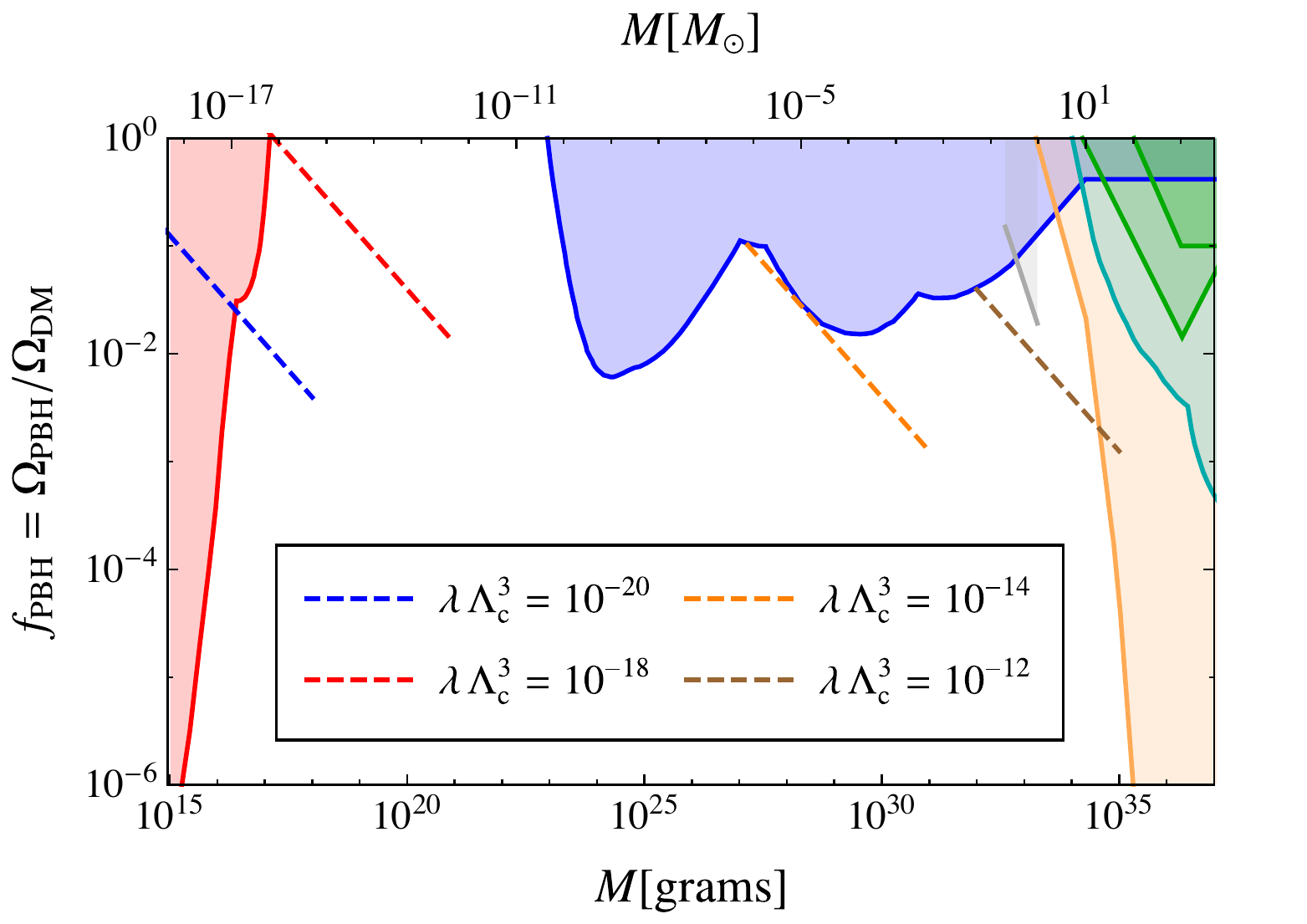}
  \caption{Mass dependence of $\drm f_{\rm PBH} / \drm \ln M_{\rm PBH}$ for different values of 
      the combination $\lambda \Lambda_{\crm}^{3}$ according to Eq.~\eqref{eq:fpbh}. 
      \textit{Shaded areas}: 
      Constraints on $f_{\rm PBH}$ for a monochromatic mass function, 
      from evaporations (red), 
      lensing (blue), 
      gravitational waves (GW) (gray), 
      dynamical effects (green), 
      accretion (light blue), 
      CMB distortions (orange) (see Ref.~\cite{Carr:2020xqk}).
      }
  \label{fig:fpbhlnm}
\end{figure}

Figure~\ref{fig:fpbhlnm} shows $\drm f_{\rm PBH} / \drm \ln M_{\rm PBH}$ for different values of the combination $\lambda \Lambda_{\crm}^{3}$; the shaded areas correspond to observational constraints on the PBH abundance (see Ref.~\cite{Carr:2020xqk}). The displayed constraints are computed for a monochromatic spectrum. Therefore, the fact that the dotted lines do not intersect the shaded areas, does not imply viability for the scenario in that mass range and should be considered only as an indication of such.

In fact, the extended spectrum found in Eq.~\eqref{eq:fpbh} scaling as $M_{\rm PBH}^{-1/2}$ is heavily constrained. A possible spectrum cut-off, is given by the tunnelling probability $P_{\rm tunnel}$ of the string into quark/anti-quark pairs per unit-time per unit-length. Indeed, as the configuration becomes longer than a certain critical length, it starts splitting into smaller configurations \cite{Vilenkin:2000jqa}. This can be quantified as
\begin{equation}
    P_{\rm tunnel}
       \propto
                    e^{- \pi\,m_{\qrm}^{2} / \Lambda_{\crm}^{2}}
                    \; ,
                    \quad
    P_{\rm tunnel}\,t_{\rm crit}^{2}
        \overset{!}=
                    1
                    \; ,
\end{equation}
where the second condition defines the critical length $t_{\rm crit}$. This effect matters for PBH mass of order $M_{\rm PBH} \sim \Lambda_{\crm}^{2}\.P_{\rm tunnel}^{-1/2}$. Justified by the assumption $\Lambda_{\crm} \ll m_{\qrm}$, we neglect in the following discussion this possibility, although this could be phenomenologically relevant for certain regions of the parameter space. Therefore, our scenario is very different from other mechanisms, where the spectrum is exponentially suppressed (\cf~Ref.~\cite{Carr:2017jsz}).

\begin{figure}[t]
  \centering
  \includegraphics[width=0.48\textwidth]{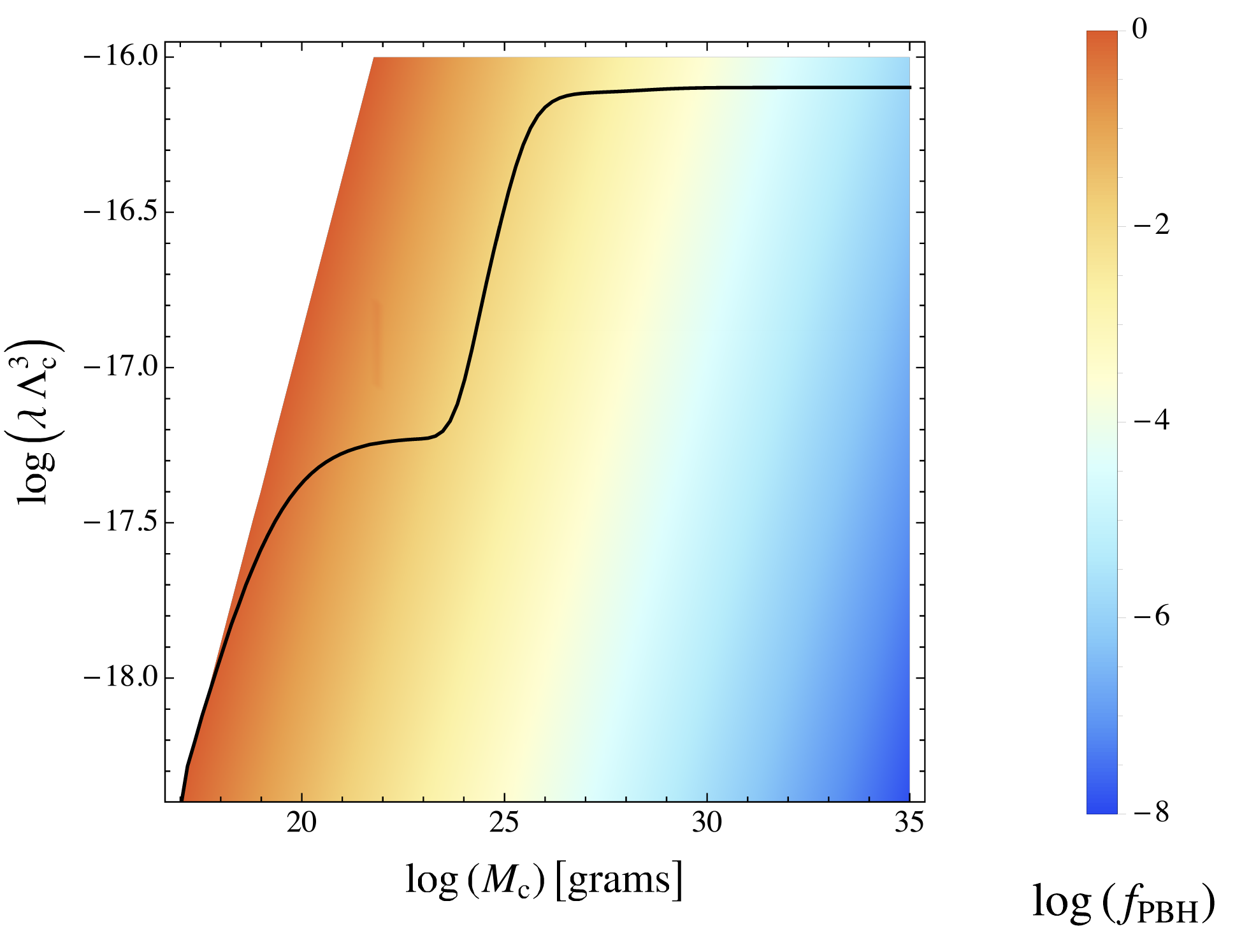}
  \caption{Dependence of $f_{\rm PBH}$ on the highest value of the mass-spectrum, $M_{\crm}$,
      and the combination $\lambda \Lambda_{\crm}^{3}$ [see~Eq.~\eqref{eq:fpbh}].
      The black line corresponds to saturation of Eq.~\eqref{eq:conditionext} 
      below which the scenario admits PBH as dark matter.
      }
  \label{fig:fpbhext}
\end{figure}

In order to check whether it is possible to accommodate all of the dark matter with the confinement mechanism, we perform a constraint analysis taking the extended nature of the PBH mass function into account. Therefore we impose the condition \cite{Carr:2017jsz}
\begin{equation}
\label{eq:conditionext}
    \int_{M_{1}}^{M_{2}}\drm \ln M_{\rm PBH}\;
    \frac{ \drm f_{\rm PBH}( M_{\rm PBH} ) }
    {\drm \ln M_{\rm PBH} }
    \frac{ 1 }{ f_{\rm max}( M_{\rm PBH} ) }
        \overset{!}{\leq}
                    1
                    \; ,
\end{equation}
where $f_{\rm max}( M_{\rm PBH} )$ corresponds to the maximal allowed value of $f_{\rm PBH}$ in the case of a monochromatic spectrum as given in Fig.~\ref{fig:fpbhlnm}. In order to proceed with this analysis, we focus on the region between $M_{1} = 10^{17}\,\grm$, $M_{2} = 10^{37}\,\grm$ and depict our results in Fig.~\ref{fig:fpbhext}. Note that $M_{\crm}$ corresponds to the highest mass in $\drm f_{\rm PBH} / \drm \ln M_{\rm PBH}$. The black line is obtained by saturating the inequality in Eq.~\eqref{eq:conditionext}. Values below that are compatible with experimental constraints. We note that for a maximum mass $M_{\crm}$ between $10^{17}\,\grm$ and $10^{19}\,\grm$, corresponding to $\lambda \Lambda_{\crm}^{3} \approx 10^{-18}$, the entire dark matter can be accommodated for. In order to obtain a PBH of mass $10^{17}\,\grm$, the corresponding choice of $\lambda \approx 10^{-3}$ and $\Lambda_{\crm} \approx 10^{-5}$ requires $T \simeq 8 \cdot 10^{-16}$ [see Eqs.~(\ref{eq:gravmass1}, \ref{eq:ttoT})]. 

Of phenomenological relevance could also be the case where PBHs are formed during a matter-dominated epoch. This could happen if quarks are already confined within the inflationary period and form while the inflaton relaxes oscillating around its minimum at the end of inflation or if the collapse takes place after equality (therefore contributing negligibly to the dark matter). In both cases, the PBH dark-matter fraction scales as
\begin{equation}
  f_{\rm PBH}^{\rm md}
    \propto
                \ln{M_{\rm PBH}}
                \; .
\end{equation}
It follows that in the case of an extended spectrum, such as the one considered in Fig.~\ref{fig:fpbhext}, the parameter window realising order-one fraction of dark matter is smaller. This is not a problem as matter-dominated periods at the end of inflation cannot be too long and the mass spectrum will not span such a wide range of different masses. 

Note that the proposed scenario requires no exponential fine-tuning, therefore avoiding one of the biggest issues plaguing the standard mechanism for PBH production based on the collapse of horizon-size overdensities. In fact, if the PBHs form from Gaussian inhomogeneities with root-mean-square amplitude $\sigma$, then the fraction $\beta$ of horizon patches undergoing collapse to PBHs is~\cite{Carr:1975qj}
\begin{equation}
	\beta
		\approx
				{\rm erfc}\!
				\left[
					\frac{\delta_{\crm}}
					{ \sqrt{2\,} \, \sigma }
				\right]
				,
				\label{eq:beta(T)}
\end{equation}
where `erfc' is the complementary error function and $\delta_{\crm} \equiv \delta \rho_{\crm} / \rho$ corresponds to the critical threshold for black-hole formation.\footnote{During the radiation-domination, assuming spherical collapse, a value of $\delta_{\crm} = 0.45$ has been obtained using numerical simulations (\cf~Ref.~\cite{Musco:2008hv}). This value depends on the shape and the statistics of the overdensities which undergo gravitational collapse (see Refs.~\cite{Escriva:2019phb, Musco:2020jjb} for spherical perturbations and Ref.~\cite{Kuhnel:2016exn} for non-spherical shapes). Furthermore, there is a slight discrepancy amongst the results of various groups.} Equation \eqref{eq:beta(T)} makes the exponential sensitivity of the PBH abundance on the amplitude of the primordial power spectrum apparent. It is clear that the proposed model avoids this sort of exponential fine-tuning, while still allowing for a phenomenological-window capable of justifying an order-one fraction of the presently observed dark matter.

%%%%%%%%%%%%%%%%%%%%%%%%%%%%%%%%%%%%%
\subsection*{Supermassive Black Holes}

One of the open questions in current astrophysics is related to the origin of supermassive black holes (SMBH), namely those black holes whose mass exceeds approximately $M_{\rm BH} \gtrsim 10^{5}\,M_{\odot}$, located in the galactic centres \cite{2009ApJ...704.1135B}. These heavy objects are difficult to be produced by standard accretion of stellar black holes. Indeed this mechanism becomes effective at late times (\ie~at redshifts between $0 \leq z \leq 6$), while heavy black holes with masses $M_{\rm BH } \gtrsim 10^{9}\,M_{\odot}$, have been observed earlier ($z \gtrsim 6$) as discussed in Refs.~\cite{volonteri2010formation, serpico2020cmb}. This therefore suggests a different production mechanism. The observational ratio between the mass of a galaxy, $M_{\rm galaxy}$, and that of the supermassive black holes, $M_{\rm BH}$, at its centre is approximately \cite{2009ApJ...704.1135B}
\begin{equation}
    \frac{ M_{\rm{galaxy}} }{ M_{\rm BH}^{\rm centre} }
        \sim
                    10^{5}
                    \; .
\end{equation}
Combining this with the number density of galaxies massive enough to host a SMBH \cite{2016ApJ...830...83C}, an amount of PBHs of order $f_{\rm PBH}^{\rm centre} \sim 10^{-7}$ is required (see Ref.~\cite{Bernal:2017nec})\footnote{The value of $f_{\rm PBH}^{\rm centre}$ might change by several orders of magnitude due to uncertainties in the accretion mechanism ($f_{\rm PBH}^{\rm centre}$ could be as low as $10^{-9}$ as commented in Ref.~\cite{Bernal:2017nec}).}. These super-heavy objects are a natural byproduct of our mechanism. 

In fact, if we focus on the window where an order-one fraction of dark matter is produced via confinement, namely for a maximal mass $M_{\crm} \sim 10^{19}\,\grm$, $\lambda \Lambda_{\crm}^{3}\sim 10^{-18}$, we obtain
\begin{equation}
\label{eq:fpbhsmbh}
    f_{\rm PBH}^{\rm heavy}
        \equiv
                    \int_{10^{4} M_{\odot}}^{M_{\rm eq}} \drm \ln M_{\rm PBH}\;
                    \frac{\drm f_{\rm PBH}( M_{\rm PBH} ) }{ \drm\ln M_{\rm PBH} }
        \approx
                    10^{-10}
                    \; .
\end{equation}
The resulting value $f_{\rm PBH}^{\rm heavy}$ is essentially independent w.r.t.~the upper cut-off already a few orders of magnitude above the lower one. Finally, taking into account accretion, PBHs for which $M_{\rm PBH} \gtrsim 10^{4}\,M_{\odot}$ at formation, can increase their value up to the one observed of $10^{9}\,M_{\odot}$ as seen in Ref.~\cite{Pacucci:2017mcu}. This accretion mechanism, however, slows down for heavier masses $M_{\rm BH} \gtrsim 10^{10}\,M_{\odot}$ as discussed in Ref.~\cite{carr2021constraints}. This process may increase $f_{\rm PBH}^{\rm heavy}$ to the necessary value of $10^{-7}$, which would be in complete agreement with the recent analysis of Ref.~\cite{serpico2020cmb}. There it is shown, via semi-analytical arguments, that PBHs of mass $M_{\rm PBH} \gtrsim 10^{4}\,M_{\odot}$, not only can accrete to the desired SMBH masses, but they can do so without conflicting with current observational bounds, justifying the current amount of observed SMBHs as long as $f_{\rm PBH}( M \gtrsim 10^{4}\,M_{\odot} ) \lesssim 10^{-9}$. Direct comparison with Eq.~\eqref{eq:fpbhsmbh} shows that SMBHs are a natural consequence of PBHs generated via confinement.

%%%%%%%%%%%%%%%%%%%%%%%%%%%%%%%%%%%%%%%%%%%%%%%%%%%%%%%%%%%%%%
\subsection*{Evaporation Constraints}
\label{sec:Evaporation-Constraints}

The usual assumption of the validity of Hawking radiation \cite{hawking1974black, hawking1975particle} throughout the entire lifetime of a black hole leads to disregard PBHs with masses below approximately $10^{15}\,\grm$ as possible dark-matter candidates, as these would have fully evaporated until now.

However, these considerations are based on extrapolations of the semiclassical computation beyond the domain of its validity. It is evident that the quantum back-reaction due to decay, becomes significant latest by the time when a black hole emits about half of its mass.

The computations \cite{Dvali:2011aa} of the back-reaction, performed within an explicit microscopic theory of a black hole \cite{Dvali:2011aa}, fully confirm this intuition. It is clear that by its half-decay, the true quantum evolution of a black hole fully departs from the semiclassical behaviour. One important aspect of this departure is the back-reaction from the quantum information carried by a black hole \cite{Dvali:2018xpy}. 

It has been argued in Ref.~\cite{Dvali:2020wft} that, due to an inability of a black hole to get rid of this information efficiently, the Hawking radiation slows down dramatically, and may eventually even come to halt. The studies performed on various prototype models unambiguously indicate the slow-down of the decay process latest by the time of a half-decay. While understanding the fate of a black hole beyond this time requires more careful analyses, the existing studies provide enough evidence to conclude that it is premature to discard PBHs lighter than $10^{15}\,\grm$ from the list of dark matter candidates. 

Here, we shall adopt the most conservative result of Ref.~\cite{Dvali:2020wft}, namely, assuming a correction to the black-hole lifetime $\tau$ of the form
\begin{equation}
\label{eq:newlife}
    \tau
        \,\ra\,
                    S^{2}\.\tau
                    \; ,
\end{equation}
with $S$ being the BH entropy. It can be easily seen that PBHs, subject to Eq.~\eqref{eq:newlife}, of masses larger than $M_{\rm BH} \geq 10^{4}\,\grm$ have a lifetime longer than the age of the Universe, $t_{0}$, and therefore might indeed be possible dark-matter candidates. As such, they would still be constrained from BBN and CMB in the region $10^{10}\,\grm$ up to $10^{15}\,\grm$, although in a weaker way, due to a reduced intensity of the emitted radiation. For PBHs of mass $M_{\rm PBH} \geq 10^{16}\,\grm$ the semi-classical description is applicable, so there are no changes to the phenomenological constraints of $f_{\rm PBH}$.

\begin{figure}[t]
\centering
\includegraphics[width=0.48\textwidth]{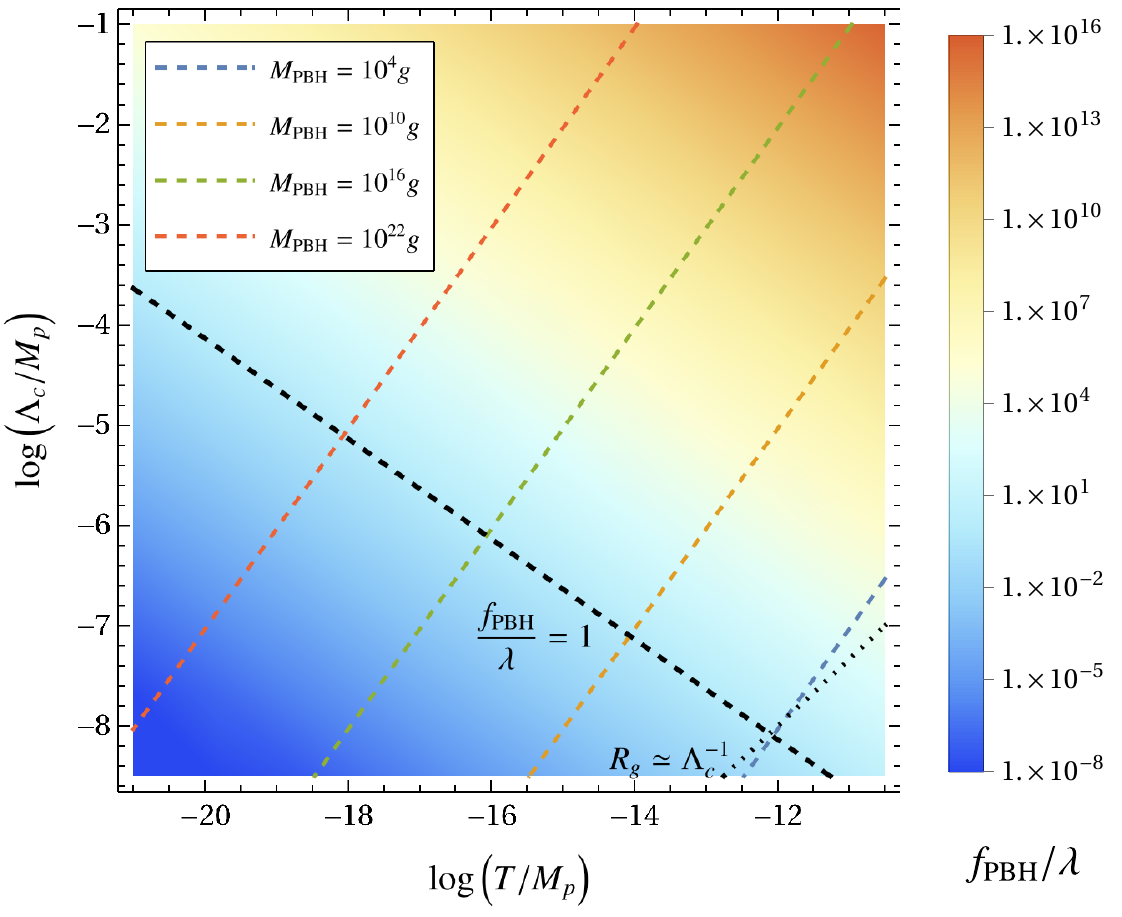}
\caption{Density plot of $f_{\rm PBH} / \lambda$ vs.~$T$ and $\Lambda_{\crm}$ 
    according to Eq.~\eqref{eq:fpbh}. 
    The dashed black line corresponds to $f_{\rm PBH} / \lambda = 1$, 
    while the dotted black line is given by $R_{\grm} \simeq \Lambda_{\crm}^{-1}$.
    }
    \label{fig:density3}
\end{figure}

In Fig.~\ref{fig:density3} we show the dependence of $f_{\rm PBH}$ for different values of the confining scale $\Lambda_{\crm}$ and temperature $T$. There is a rather big window between $10^{4}\,\grm$ and $10^{10}\,\grm$ where a $100\%$ fraction of the dark matter could  be explained by PBHs. The dotted line $R_{\grm} \simeq \Lambda_{\crm}^{-1}$ corresponds to the limiting case for black-hole formation, as in this case the gravitational radius becomes comparable to the thickness of the QCD string.
\vs{-2mm}

%%%%%%%%%%%%%%%%%%%%%%%%%%%%%%%%%%%%%%%%%%%%%%%%%%%%%%%%%%%%%%
\subsection*{Gravitational Waves}
\label{sec:GWs}

The collapse of a quark/anti-quark pair into a PBH is accompanied by emission of gravitational radiation. The process is very similar to the emission of gravity waves from the monopole/anti-monopole system connected by a cosmic string \cite{Martin:1996cp}. In our case, since the black holes are formed in a regime of a very small impact parameter, the emission of gravity waves can be estimated using the approximation of a straight QCD-string attached to quarks.

The emitted frequencies due to the collapse of a single pair start (in the infrared) from approximately the inverse initial distance $t$. Since the impact parameter matters on distances comparable to the configuration's Schwarzschild radius, and we are interested here in the spectrum produced at much lower frequencies, the dynamics can be assumed to be head-on; the results of Ref.~\cite{Martin:1996cp} can be applied, leading to an emitted gravitational-wave power $P_{n}$ at frequency $\omega_{n} = 2 \pi n / t$ of
\begin{equation}
\label{eq:pn}
	P_{n}
	    \sim
	                \frac{ \Lambda_{\crm}^{4} }{ n }
	                \; .
\end{equation}
Equation~\eqref{eq:pn} is valid up to frequency numbers of order $M_{\rm PBH}^{2} / m_{\qrm}^{2}$. Above, $P_{n}$ decreases first as $n^{-2}$, and is eventually exponentially suppressed \cite{Martin:1996cp, Martin:1996ea, Babichev:2004gy}. For practical purposes, Eq.~$\eqref{eq:pn}$ can be safely used through our discussion. Moreover, note that the power \eqref{eq:pn} is emitted (at formation time) in a small beaming angle oriented towards the direction of collapse; it can be estimated as $\theta \sim n^{-1/2}$ \cite{Martin:1996cp}. Therefore, taking into account that many pairs collapse into PBHs through the cosmological history, it is possible, for some range of the parameter space, to obtain an observationally relevant stochastic gravitational-wave background.

A similar analysis was performed in Ref.~\cite{leblond2009gravitational,buchmuller2021stochastic}, where the gravitational-wave signal due to the rapid annihilation of monopole pairs, linked by straight strings, was studied, and it was found to lead to a scale-invariant spectrum for $\Omega_{\rm GW}$. Similarly, also for the case studied in our work, assuming the mass distribution extends to $M_{\rm PBH} \gg \Lambda_{\crm}^{2}\,f^{-1}$, $f$ being the gravitational-wave frequency, the resulting spectrum is nearly scale invariant.

For example, the recently observed NANOGrav data, found at a frequency of about $f = {\rm{year}}^{-1}$, are mapped to a formation time within the confinement scenario of around $10^{-3}\,{\srm}$. The stochastic gravitational-wave background can easily be estimated following Ref.~\cite{leblond2009gravitational, damour2001gravitational, buchmuller2021stochastic} giving
\begin{equation}
	\Omega_{\rm GW}^{\rm NANO}
        \sim
	                10^{-11}
        \propto
                     \lambda \Lambda_{\crm}^{4}
                    \; ,
                    \label{eq:nano}
\end{equation}
where the prefactor depends on the extension of the distribution of PBH spectrum, and is of $\Ocal( 1 )$ if PBHs are still forming today. It follows that $\Lambda_{\crm} \geq 10^{-2.5}$, compatible with the fact that a low confinement scale, suppresses the amount of emitted gravitational waves. Moreover, the amplitude \eqref{eq:nano} required by the NANOGrav data, combined with the flat spectrum generated by the confinement mechanism, lies within the prospective reach of LISA \cite{buchmuller2021stochastic}. 

We would like to point out that a mild extension of our scenario can provide a significant (or even a dominant)  scalar component of gravity waves. This would be the case if there would exist a massless scalar field interacting both with heavy quarks as well as with the Standard Model particles. For example, it suffices to couple the mediator scalar to the heavy quarks via a standard Yukawa interaction with coupling constant $g$.
 
In order to have an interesting observable effect, this coupling must be stronger than gravity. Let us parameterise by $\epsilon$ the relative strength of the effective coupling of the scalar to the Standard Model particles with respect to gravity. This quantity must be very small in order to accommodate phenomenological constraints on gravity-competing forces. In particular, the equivalence-principle violating part of such interaction must obey $\epsilon \lesssim 10^{-6}$ \cite{touboul2017microscope, berge2018microscope}. In case of an equivalence-preserving interaction, the constraints are milder, $\epsilon \lesssim 10^{-3}$ or so \cite{blasone2018equivalence}.  

Under such circumstances the power of the scalar gravitational radiation relative to a tensor one, as seen by a detector constructed out of ordinary Standard Model matter, is controlled by the product $g \epsilon$. At the expense of having large $g$, this parameter can be within the range of observational interest, even for the values of $\epsilon$ well within the above phenomenological bounds. This provides an interesting motivation for the search of gravitational waves with rather unusual properties. In particular, such waves can carry a scalar monopole component and their observed intensity can be sensitive to an isotope composition of the detector.  

The possible existence of a scalar component of gravity waves can also be of potential interest in the light of the NANOGrav data (see Ref.~\cite{NANOGrav:2020bcs}) which allows for the possibility that the observed gravitational-wave signal receives contributions also from monopolar and/or dipolar components. This is impossible in General Relativity, where gravitational waves require a quadrupolar source \cite{1983ApJ...265L..39H}. However, as discussed above, the confinement mechanism can in principle accommodate additional non-standard contributions into gravitational waves.

%%%%%%%%%%%%%%%%%%%%%%%%%%%%%%%%%%%%%%%%%%%%%%%%%%%%%%%%%%%%%%
\subsection*{PBHs from Real QCD?}
\label{sec:PBHs-from-QCD}

In this Section we wish to briefly discuss whether the ``real" QCD can be used for realising the presented PBH scenario. At first glance, this looks improbable, since ordinary QCD does not satisfy the necessary condition of all quarks being heavier than the QCD scale $\Lambda_{\crm}$. Correspondingly, one expects that strings have no chance to last any appreciable time. Even if formed, they will quickly break apart by producing a multiplicity of light quark pairs. However, the situation is more subtle, because for our mechanism the present-epoch values of the quark masses relative to $\Lambda_{\crm}$ are not important. What matters is their values in the early Universe.

As argued in Ref.~\cite{dvali1995removing}, the quark masses as well as $\Lambda_{\crm}$ can significantly vary during cosmological evolution. The reason is that the expectation values of the fields controlling these parameters change. In fact, in string-theoretic context, where the QCD gauge coupling ``constant" is set by fields, such as dilaton and other moduli, the time-variation of $\Lambda_{\crm}$ is a norm rather than an exception. The same is true about the quark masses, since both the Higgs expectation value and Yukawa couplings (which similarly to gauge coupling are determined by moduli) vary. 
 
Therefore, when submerged into a cosmological framework, ordinary QCD can satisfy all requirements of PBH formation scenario by quark confinement. In the present paper we shall not enter into the details of model building, since our purpose was to point out rather generic aspects of the scenario. However, we wish to remark that its specific realisations can lead to additional potentially interesting correlations with phenomena that are sensitive to $\Lambda_{\crm}$ and $m_{\qrm}$. An example is an expansion of the cosmologically acceptable window for the axion field, due to variations of $\Lambda_{\crm}$ and $m_{\qrm}$ in the early epoch \cite{dvali1995removing}.

Due to the electromagnetic charge of QCD quarks, a question may arise whether the electromagnetic backreaction can affect the collapse dynamics. As shown in Ref.~\cite{Berezinsky:1997kd}, the power in photon emission is $P\sim \Lambda_{\crm}^4/ m_{\qrm}^{2}$. This is negligible w.r.t.~the initial total energy of the system under condition \eqref{eq:assumptscales} which is assumed throughout this work. Furthermore, as previously mentioned, for the scenario to work with ordinary QCD, the parameter values, such as QCD scale and quark masses must be shifted from their present values. In this case, there is no reason why also the electromagnetic coupling could not have been much weaker during the collapse, therefore suppressing the effect even further.
\vs{-3mm}

%%%%%%%%%%%%%%%%%%%%%%%%%%%%%%%%%%%%%%%%%%%%%%%%%%%%%%%%%%%%%%
\subsection*{String-Theoretic Realisation}

We wish to briefly discuss a possible string-theoretic realisation of our scenario. We shall not give a full-fledged construction, but merely only point out that string theory contains all the generic key ingredients for it in form of $D_{p}$-branes. These represent extended objects, of world-volume dimensionality $p + 1$, on which open strings end (for introduction, see Ref.~\cite{Polchinski:1998rr}).

It has been known for some time that $D$-branes can serve as the main engine for driving inflation in string theory \cite{dvali1999brane, dvali105203d}. In this scenario, the $D$-brane tension serves as the source for the energy density required for achieving a temporary de Sitter-like state. In the simplest version, initially, some $D$-branes and anti-$D$-branes ($\bar{D}$-branes) are  separated by finite distances in compact dimensions, and slowly move towards each other. We note that since $D$-branes can be wrapped around some of the compact dimensions, the  dimensionality of the transverse space in which they are separated can be less than $6$. The positive potential energy provided by this configuration (approximately given by the sum of the brane tensions) drives inflation along the $4$ non-compact dimensions. The r{\^o}le of a slow-rolling inflaton field is played by a brane-separation mode. Upon collision, the $D$-$\bar{D}$ pairs annihilate, releasing their tension energy into various low-lying string excitations, thereby reheating the Universe.
 
It was noticed that brane inflation allows production of extended objects much heavier than the inflationary Hubble parameter $H_{\irm}$, such as, macroscopic fundamental strings \cite{dvali1999infrared}. In the present paper we shall focus on the production of stringy objects as a result of brane annihilation after inflation. During the process of annihilation of $D_{p}$-$\bar{D}_{p}$, the $D_{p - 2}$-branes are generically produced. The structures that have only two world-volume coordinates in ``our" $4$ non-compact space-time dimensions, from the point of view of effective $4$-dimensional theory, represent $D$-strings. The characteristic tension of such strings is given by $\mu \sim M_{\srm}^{2} ( M_{\srm}^{n}\.V_{n} ) / g_{\srm}$, where $M_{\srm}$ is the fundamental string scale, $g_{\srm}$ is the string coupling and $n$ is the the number of compact longitudinal dimensions (wrapped by the $D$-brane). $V_{n}$ is the volume of these dimensions. In the simplest case, if the compactification volume is given by the string length, we have $( M_{\srm}^{n}\.V_{n} ) \sim 1$. Remarkably, as it was argued in Ref.~\cite{sarangi2002cosmic}, after brane inflation, the production is dominated by above string-like objects. Various aspects of their formation and evolution has been studied in Refs.~\cite{copeland2004cosmic, dvali2004formation}. 

We envisage the following implementation of our PBH scenario in this framework. The lower-dimension $D$-branes, which appear point-like from the $4$-dimensional perspective, can be produced as a result of de Sitter fluctuations during the inflationary stage, while the inflation-driving $D$-$\bar{D}$ system is still functioning. From the point of view of our scenario, such point-like $D$-defects, play the r{\^o}le of ``quarks". Their mass can be estimated as $m_{\qrm} \sim M_{\srm}\.( M_{\srm}^{n}\.V_{n} ) / g_{\srm}$, where $n$ now has to be understood as the number of compact longitudinal dimensions wrapped by a ``$D$-quark", and $V_{n}$ as the volume of these dimensions. During the stage of brane-inflation, ``$D$-quarks'' are inflated away. 

Towards the end of inflation, upon annihilation of the ``parental'' $D$-$\bar{D}$ system, the string-like $D$-branes will be produced. We are interested in the case in which they connect the previously produced $D$-quarks. In such a situation, $D$-strings play the same r{\^o}le as the QCD flux tube played in the confinement case. Our scenario of PBH formation thus should follow. 

In this context, it is intriguing to point out a potential connection with gravitational-wave signals observed by NANOGrav (see Ref.~\cite{Domenech:2020ers} for a PBH explanation and Ref.~\cite{NANOGrav:2020bcs} for the original reference). As already discussed in detail in the Sec.~\ref{sec:GWs}, our scenario would reach the level of this signal for relatively large values of the string tension. These values are reached for a rather conservative choice of string-theoretic parameters. For example, in case of an elementary $D_{1}$-string, the parameters are \footnote{For effective $D_{1}$-strings that are obtained by wrapping the higher extend ($p > 1$) $D_{p}$-branes around $p - 1$ compact extra dimensions, the same effective tension can be obtained 
for smaller value of $M_{\srm}$.} 
\begin{equation}
	\label{eq:parameters}
    M_{\srm}
        \sim
                    10^{17-16}\,{\rm GeV},\;g_{\srm}
        \sim
                    10^{-2}
                    \; .
\end{equation} 
This appears to be an excellent regime in which the gravitational-wave signal originates from formation of PBHs as dark-matter candidates by our mechanism. For even larger values of the tension, the dissipation in gravitational waves is significant but effective field-theoretic estimates become less reliable.

In the context of a string-theoretic implementation of our mechanism, we would like to comment on the possible presence of non-standard (monopolar and/or dipolar) components as allowed by the NANOGrav data \cite{NANOGrav:2020bcs}. As already discussed in Sec.~\ref{sec:GWs}, the presented confinement scenario can potentially accommodate these contributions in the presence of light scalar fields sourced by the endpoints of strings (i.e., heavy quarks or $D$-``quarks") and also interacting with the Standard Model particles. Interestingly, in the string-theoretic realisation of our scenario such fields are generic, as there exist scalar moduli that are sourced by the endpoint $D$-branes. If some of these fields remain sufficiently light or massless, they can be emitted by the $D$-string/quark systems in modes including the lower multiples. 

Of course, as already explained, the same scalar fields are subjected to severe constraints from precision gravitational physics. The reason is that they mediate gravity-competing long-range forces. We cam distinguish two cases. 

The first is when the scalar force in question does not violate the equivalence principle (at least, in the sector of the Standard Model particles). This is the case if the scalar is sourced by the trace of the energy momentum tensor $T^{\mu}_{\phantom{\mu}\mu}$, which is for instance the situation for the modulus that corresponds to the fluctuations of the entire compactification volume of extra-dimensional space. The phenomenological restriction is that the strength of the force mediated by such a field among ordinary particles must approximately be $10^{-5}$  times weaker than gravity \cite{will2004testing}. The restriction comes mainly from the deflection of star light by the sun.
 
The second case is when the scalar field in question is not sourced strictly by the trace $T^{\mu}_{\phantom{\mu}\mu}$ and has also other couplings. This is for instance the case with the string dilaton. In this case, the couplings to ordinary matter fields are more severely constrained, since the force mediated by such a field violates the equivalence principle. In case of a dilaton, this violation can be understood from the fact that dilaton sets the values of gauge and gravitational couplings. Due to this, it couples with different strengths to protons and neutrons and correspondingly mediates the isotope-dependent forces among the atoms (see, \eg, Refs.~\cite{Antoniadis:1997zg, Dvali:2001dd}). The phenomenological constraints on such forces bound them to $10^{-12}$ of gravity \cite{touboul2017microscope, berge2018microscope}.
 
In conclusion, while the string-theoretic version of our scenario contains interesting candidate sources for low-multipole gravitational waves, a non-trivial interplay among the couplings with various sources is required for accommodating the phenomenological constraints from the existing precision gravitational measurements.\footnote{In the present paper we shall not enter in a discussion about naturalness of light scalar moduli, as this question is not specific to the present case and is shared by any theory of scalar gravity.} 

Shall the future data be proven to indeed provide the evidence for such gravitational waves, this will boost motivation for finding viable enhancement mechanisms of the mentioned scalar waves.

%%%%%%%%%%%%%%%%%%%%%%%%%%%%%%%%%%%%%%%%%%%%%%%%%%%%%%%%%%%%%%
\section{Discussion and Outlook}
\label{sec:Discussion-and-Outlook}

In this work we have introduced a novel mechanism for PBH formation, which relies on the production of quarks during the inflationary phase. These are exponentially pushed apart, and get connected by QCD flux tubes after the temperature drops below the confinement scale. Upon horizon re-entry, the quarks accelerate towards each other under the influence of the string tension. The black-hole formation is then triggered by the large amount of energy stored in the string connecting the quark pair.

The presented mechanism does not suffer from exponential fine-tuning, unlike most standard scenarios in which PBHs form from collapse of cosmological overdensities. Moreover, unlike other PBHs which are created during the radiation-dominated epoch and which carry none to very little spin, the confinement mechanism allows for the formation of maximally-rotating black holes, in particular, in the sub-solar mass range as shown in Fig.~\ref{fig:spin}. At formation time of a black hole, the impact parameter between the colliding quarks can be as large as the configuration's Schwarzschild radius, therefore leading to maximally-rotating black holes.

In Sec.~\ref{sec:Dark-Matter} we demonstrated that the PBHs formed by the confinement mechanism can account for the entirety of the dark matter in a mass range around $10^{17}$ -- $10^{19}\,\grm$ (see Fig.~\ref{fig:fpbhext}). As a by-product, the slowly decaying mass spectrum, scaling as $M_{\rm PBH}^{-1/2}$, could {\it at the same time} provide seeds for the supermassive black holes observed in the galactic centres. In fact, the right amount of seeds is compatible with $100\%$ of PBH dark matter produced by the same mechanism. 

In view of recent arguments that imply a substantial relaxation of the evaporation constraints [see Ref.~\cite{Dvali:2020wft} and Eq.~\eqref{eq:newlife}], we found that even conservative estimates allow for possible realisations of $100\%$ of the dark matter within our scenario in the mass range $10^{4}$ -- $10^{10}\,\grm$. For masses above the upper end of this range, the standard radiative constraints due to nucleosynthesis, and CMB{\;---\;}although shallower{\;---\;}still applies when using our conservative estimate. However, a complete reevaluation of those constraints, taking the black hole's full quantum structure and dynamics into account, is still missing. This would be of great importance for any scenario in which PBHs of masses below approximately $10^{17}\,\grm$ form. We leave these investigations for future work. 

Interestingly, the confinement mechanism of PBH formation could in principle have been driven by ordinary QCD. This is possible, since the time-variation of parameters such as the QCD scale and quark masses are rather generic in inflationary cosmology \cite{dvali1995removing}. Correspondingly, it is not unnatural that the values favourable for the presented mechanism of PBH formation were attained in the early epoch.

We have also outlined a possible implementation of the presented mechanism within a string-theoretic framework of inflation driven by $D$-branes \cite{dvali1999brane, dvali105203d}. In this case, the r{\^o}le of heavy ``quarks'' connected by colour flux tubes is assumed by compact $D$-branes connected by $D$-strings. Interestingly, for conservative values of the string-theoretic parameters \eqref{eq:parameters}, the obtained gravitational-wave signal from PBH formation has the right amplitude in order to be compatible with the events recently detected by NANOGrav, including the possibility to account for possible scalar contributions to the signal.

%%%%%%%%%%%%%%%%%%%%%%%%%%%%%%%%%%%%%%%
\section*{Acknowledgements}
We are indebted to Dominik Schwarz for stimulating remarks on the possible non-quadrupolar nature of the NANOGrav signal. Discussions with Lasha Berezhiani are also acknowledged. This work was supported in part by the Humboldt Foundation under Humboldt Professorship Award, by the Deutsche Forschungsgemeinschaft (DFG, German Research Foundation) under Germany's Excellence Strategy - EXC-2111 - 390814868, and Germany's Excellence Strategy under Excellence Cluster Origins.

%%%%%%%%%%%%%%%%%%%%%%%%%%%%%%%%%%%%%%%%%%%%%%%%%%%%%%%%%%%%%%
\setlength{\bibsep}{4pt}
\bibliography{citations}
\end{document}